\documentclass[journal=jpclcd,manuscript=article,doi]{achemso}
\setkeys{acs}{doi = true}

\usepackage{graphicx}
\usepackage{lmodern} 
 \usepackage{amsmath} \usepackage[english,american]{babel}
\usepackage{makecell}

\newcommand{\pathfig}{.//}

\title{Superlattice Induced by Charge Order in the Organic Spin Chain (TMTTF)$_2X$ ($X=$  SbF$_6$, AsF$_6$ and PF$_6$) Revealed
	by High Field EPR}

\author{Charles-Emmanuel Dutoit} \affiliation[1]{Aix-Marseille Universit\'{e}, CNRS, IM2NP (UMR
7334), Marseille, France}

\author{Anatoli Stepanov} \affiliation[1]{Aix-Marseille Universit\'{e}, CNRS, IM2NP (UMR
7334), Marseille, France}

\author{Johan van Tol} \affiliation[2]{The National High Magnetic Field Laboratory,
Florida State University, Tallahassee, Florida 32310, USA}

\author{Maylis Orio} \affiliation[3]{Aix-Marseille Universit\'{e}, CNRS, ISm2, Central Marseille, Marseille, France}

\author{Sylvain Bertaina} \email{sylvain.bertaina@im2np.fr}
\affiliation[1]{Aix-Marseille Universit\'{e}, CNRS, IM2NP (UMR 7334), Marseille,
France}

\begin{document}

\begin{tocentry}
%
		\includegraphics[width=\linewidth,bb=40 40 570 500,clip]{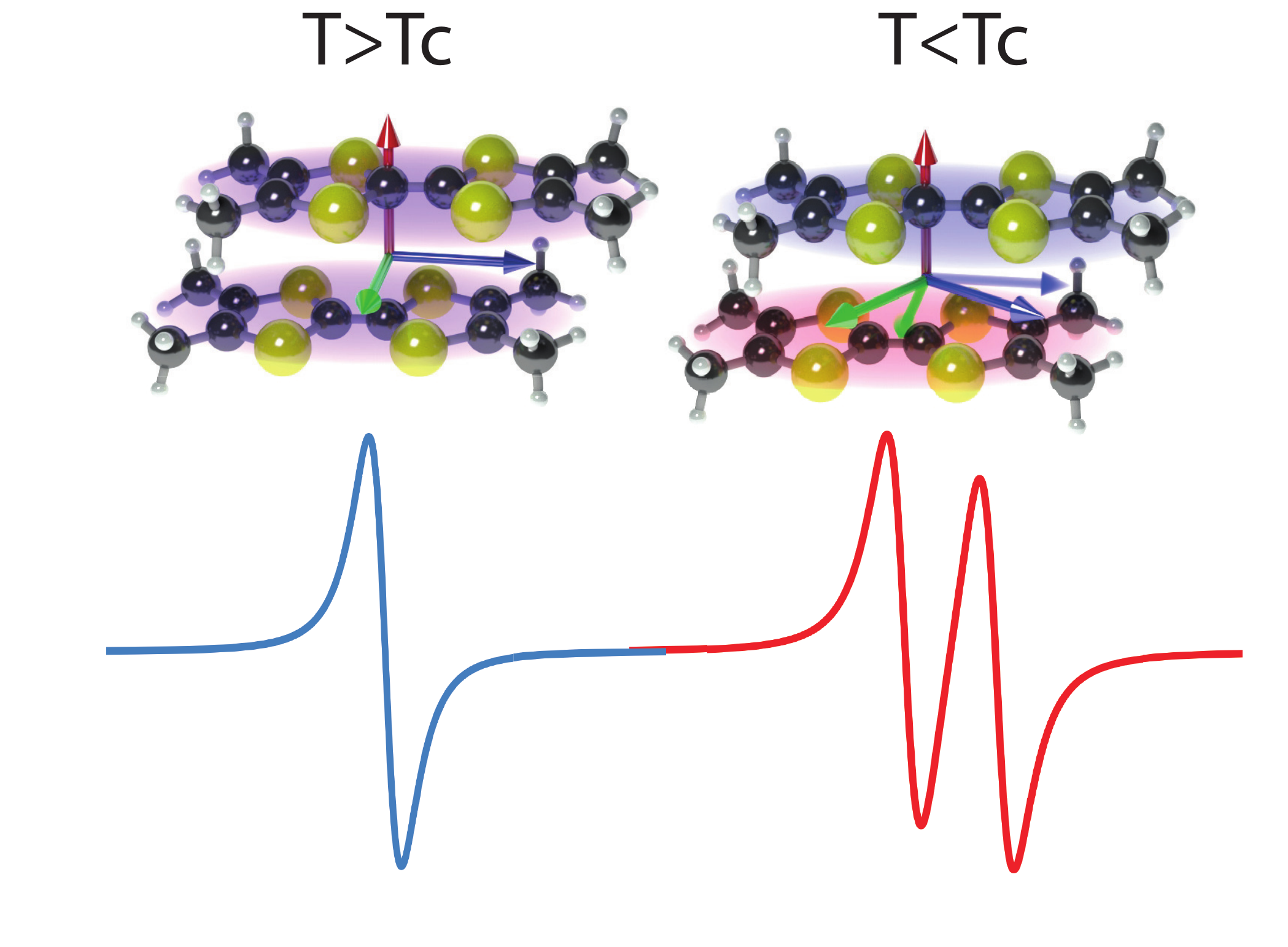}
%
\end{tocentry}

\begin{abstract} 
We have investigated the charge ordering phase of the quasi one dimensional
quantum antiferromagnet (TMTTF)$_2X$ ($X=$  SbF$_6$, AsF$_6$ and PF$_6$) using
high fields/frequencies electron paramagnetic resonance. In addition to the
uniform displacement of the counter anions involved in the charge order phase,
we report the existence of a superlattice between the spin chains in the
direction $c$, caused by the space modulation of the charge order. When the field is high enough,
 the magnetic decoupling of the spin chains allows us to estimate
the interaction between the chains, $J_c<1$~mK, three orders of magnitude lower
than expected from the mean field theory.
\end{abstract}

\maketitle

\section{Introduction}

The family of quasi-one-dimensional organic conductors (TMTTF)$_2X$ is known to
have a rich phase diagram with a sequence of competing ground states between
spin-Peierls (SP), antiferromagnetic (AF) or superconductor state, depending on
the nature of counter anion X or external pressure.  More particularly, the
centro-symmetric $X$ ($X=$ SbF$_6$, AsF$_6$ and PF$_6$) is metallic at high
temperatures, then it is an insulator, and finally, the ground state is either AF for SbF$_6$ ($T_N=8$~K) or SP
for AsF$_6$ ($T_{SP}=19$K) and PF$_6$ ($T_{SP}=13$~K). The
metal-insulator transition was first observed by conductivity
\cite{Laversanne1984} and microwave measurements  \cite{Javadi1988}, and was
attributed to a charge ordering (CO) \cite{Shibata2001,Riera2001} inside the
molecule. At $T>T_{CO}$, the charge (hole) is equality  divided between the two
TMTTF molecules, then at $T<T_{CO}$ a displacement of charge from one TMTTF to the other
induces a $4k_f$ ordering in the direction of the intrastack $a$ axis (spin
chain axis). The CO transition was considered \textit{structureless} because no
observation of the superlattice reflection was reported \cite{Laversanne1984}.
Since it has been proven that the X-ray was responsible for the destruction of
the CO transition \cite{Foury-Leylekian2004,Coulon2015}. Only neutron scattering was able
to directly show the displacement of the lattice during the CO transition
\cite{Foury-Leylekian2010}.

	\begin{figure} \centering \includegraphics[width=0.8\columnwidth]{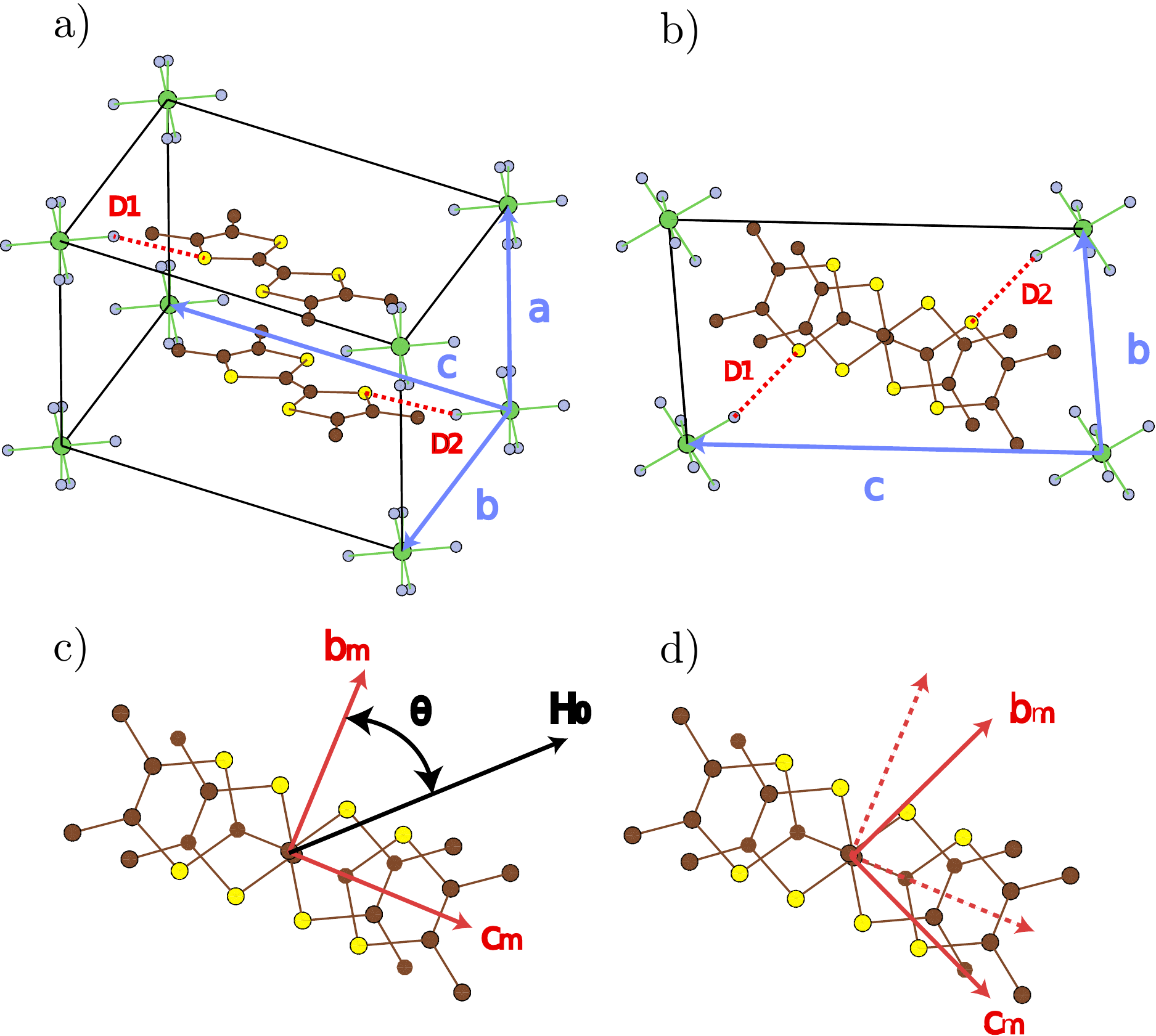}
	\caption{(color online). Crystallographic structure of (TMTTF)$_2X$. Two molecules of TMTTF (in the center of the cell) share one hole given by one of the eight counter anions $X^-$  ($X=$SbF$_6$, AsF$_6$ or PF$_6$).  } \label{fig:figstruct}
\end{figure}

The figure \ref{fig:figstruct} shows the elementary cell structure of
(TMTTF)$_2X$. The electronic properties of this family of compounds are due to a
hole shared by the two molecules of TMTTF. At high temperature ($T>T_{CO}$), the
crystal structure is triclinic with a center of symmetry P\={1}. The distances
between the sulfur atoms and the nearest counter-anion, are equal, $D_1=D_2$, 
and the density of hole is the same on each TMTTF\cite{Pouget1996}. Below
$T_{CO}$, the center of symmetry is removed, the distances $D_1$ and $D_2$ are
different, leading to a charge disproportion between the two molecules of TMTTF
and, as a consequence, to the charge ordering.

The observation of the CO transition in (TMTTF)$_2X$ has been reported using
many techniques. $^{13}$C NMR investigations showed a splitting of lines below
$T_{CO}$, caused by charge rearrangement around $^{13}$C 
\cite{Chow2000,Zamborszky2002}. A minimum of the dielectric permittivity at
$T_{CO}=156$~K for $X=$SbF$_6$, $T_{CO}=103$~K for $X=$AsF$_6$  and
$T_{CO}=67$~K for $X=$PF$_6$ have been reported by Monceau \textit{et al.}
\cite{Monceau2001} and the role of the lattice in the CO has been studied by optic
\cite{Dumm2006} and dilatometry \cite{DeSouza2008,DeSouza2010} measurements. 

Electron paramagnetic resonance (EPR) studies were also conducted. From the
magnetic point of view, the hole carries an electronic spin $S=1/2$. Due to the
low symmetry of the structure, the principal axes of the g-tensor and the
crystallographic axes are different. Let's name the crystal axes $a$, $b$ and
$c$, and the magnetic axes $a_m$, $b_m$ and $c_m$.  $a_m$, $b_m$ and $c_m$ correspond to the minimum, intermediate and maximum g factor, respectively. EPR is a tool of choice to
measure with a high accuracy the g-tensor in such organic crystals where the
anisotropy of g-factor is very weak\cite{Coulon2004}. In the high temperature
phase, the magnetic axes$c_m$ and $b_m$  are, respectively, parallel and perpendicular
to the axis of the TMTTF molecule (Fig. \ref{fig:figstruct}c) and are clearly
different to the crystallographic axes. The $a_m$ axis is perpendicular to the plane
formed by the TMTTF and is close to the $a$ axis, making with it an angle of about 3$^\circ$.
Despite a weak magnetoelectric coupling, the EPR was able to detect the CO
transition.  Conductive EPR has shown a change of line asymmetry at
$T_{CO}$\cite{Coulon2007}. A new source of line broadening below $T_{CO}$ was
reported \cite{Furukawa2005,Nakamura2003a,Nakamura2003}. On the basis of molecular
density-functional theory (DFT) calculations, the rotation of the principal axes
of the g-tensor was attributed to the CO transition \cite{Dutoit2015}. This last
result is represented in figure \ref{fig:figstruct}c and d.

It is important to notice that all these results have assumed or shown a uniform
displacement of the anions.

The (TMTTF)$_2X$ family is also considered as a good prototype of quantum
quasi-1D antiferromagnetic Heisenberg system (QQ1DAFH) with $J\sim400$~K
\cite{Dumm2000,Foury-Leylekian2009,Coulon2004} but the magnetic dimensionality
remains controversial. Band structure calculations have shown a transfer integral
inside the chain of $t_a\sim200$~meV and between the chains $t_b\sim40$~meV
\cite{Granier1988,Giovannetti2012}  and $t_c\sim1$~meV \cite{Grant1983}  in
the directions $b$ and $c$, respectively. From a simple tight binding model
\cite{Seo2006} it follows that $J_a> J_b\gg J_c$ (Using $J_a\sim
	400$~K, we estimate $J_b \sim 16$~K  and $J_c\sim 0.01$~K ). But recently
\cite{Yoshimi2012}, it has been shown that such values of the transfer integrals
could lead to a more $2D$ magnetic behavior and the application of QQ1DAFH
models \cite{Yasuda2005} might be wrong. Moreover, on account of the
quasi-absence of electronic correlation in the $c$ direction, the majority of
theoretical and experimental studies probe the properties in the $ab$ plane only.

In this article, we report a direct observation of superlattice in the
direction $c$, induced by the CO transition. Using high field EPR, we demonstrate
that, in addition to the superlattice inside the chain axis, the suppression of
the center of symmetry also creates a superlattice between the chains which is
resolved when the magnetic field is large enough to decouple the chains in the
direction $c$. The decoupling of the chains allows us to estimate the coupling
constant $J_c$. 

\section{Experimental details}

High-field/high-frequency EPR (HF-EPR) experiments have been carried out
using a homemade quasioptical superheterodyn setup developed at NHMFL
\cite{VanTol2005}. The spectrometer operates at 120~GHz, 240~GHz and 336~GHz and
at temperature from RT down to 2~K. The inhomogeneity of the field (crucial in our
results) is less than 0.1~G across the volume of the samples. The presence of
modulation coils allows us to record the first derivative of both the absorption
and dispersion signals. The single crystals of (TMTTF)$_2X$ with $X=$ SbF$_6$,
AsF$_6$ and PF$_6$ have the shape of a needle with a typical size of
$50\times100\times700~\mu$m$^3$,  small enough to avoid polariton reflections
inside the sample. The angular dependence of the EPR was measured by using a goniometer
which rotates the sample in the plane perpendicular to the $a$ axis. Due to the
triclinic symmetry, the magnetic and crystallographic axes are different. The
magnetic axes $b_m$ and $c_m$ correspond to the minimum and maximum of the
resonance field respectively. The temperature dependence of the EPR has been
measured by applying the field at 45$^\circ$ between $b_m$ and $c_m$. The HF-EPR
spectra were recorded for 	the three systems and for the three available
frequencies.

\begin{figure} \centering \includegraphics[width=0.7\columnwidth]{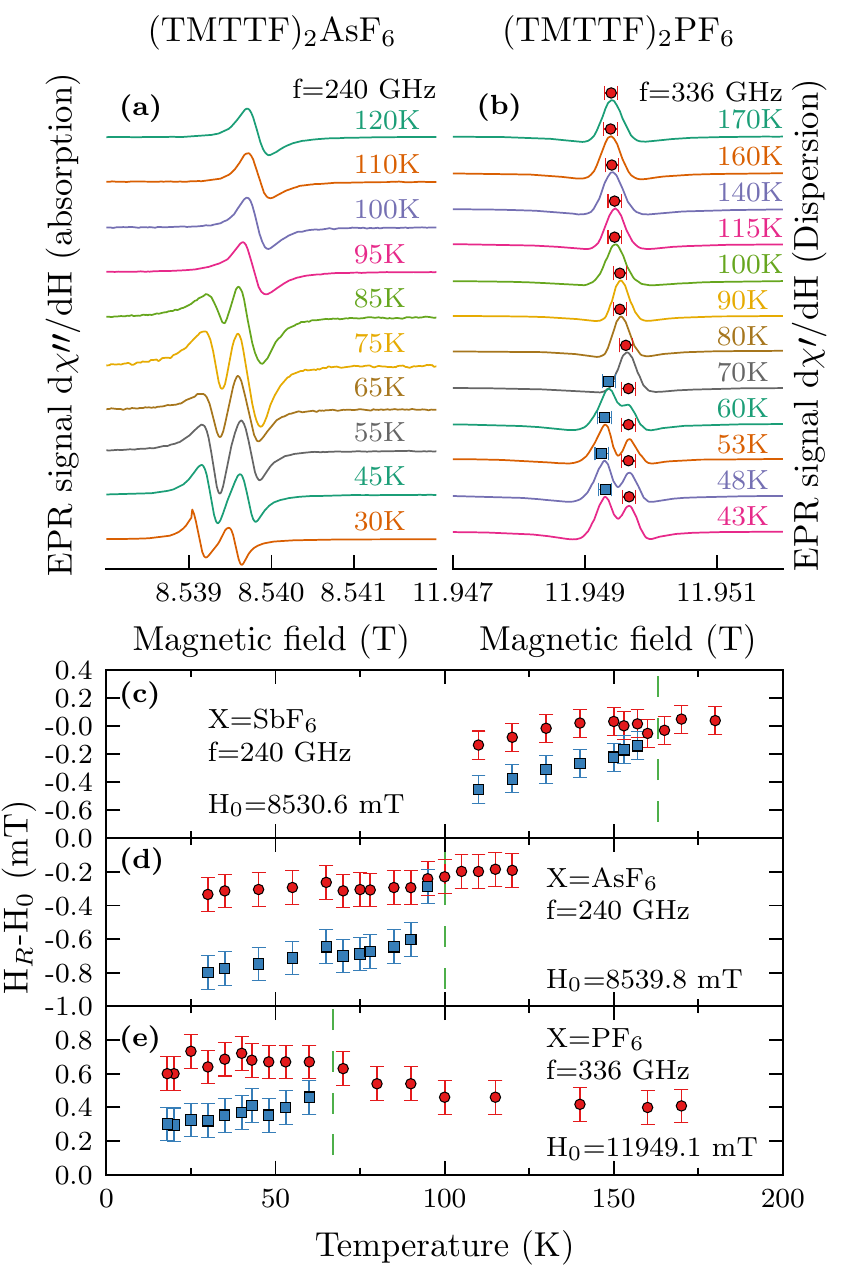}
	\caption{(color online) HF-EPR recorded at temperature above and below $T_{CO}$
		for the magnetic field at 45$^\circ$ between the magnetic axis $b_m$ and $c_m$.
		(a) Derivative absorption signals of (TMTTF)$_2$AsF$_6$ at f=240~GHz when the
		temperature decreases.(b) Derivative dispersion signals of (TMTTF)$_2$PF$_6$ at
		f=336~GHz when the temperature decreases. The red circle and blue square are
		the resonance fields $H_R$ and are reported as a function of temperature in
		(c), (d), (e) for $X=$ SbF$_6$, AsF$_6$ and PF$_6$ respectively. The vertical
		dashed lines represent $T_{CO}$. For clarity the variation of $H_R$ from the
		resonance field at high temperature $H_0$ is presented. } \label{fig:figserie}
\end{figure}

\section{Results and discussions}

\subsection{Charge order transition observed by HF-EPR}

The figure \ref{fig:figserie} illustrates the detection of the CO transition by
HF-EPR. At high temperature, the EPR line is a Lorentzian with a linewidth of
about 3~G close to the one reported at low frequency \cite{Nakamura2003}.  When
$T<T_{CO}$, the line splits in two lines of nearly the same intensity. Examples
of raw spectra are given in Fig. \ref{fig:figserie}.a and
Fig.\ref{fig:figserie}.b for $X=$AsF$_6$ and PF$_6$ respectively. For AsF$_6$
the splitting was observed at all available frequencies, while for SbF$_6$ it
was observed at 240~GHz and 336~GHz, and for PF$_6$  at 336~GHz only. The
resonance fields of the compounds are reported in Fig.\ref{fig:figserie}.c, 
Fig.\ref{fig:figserie}.d and Fig.\ref{fig:figserie}.e. The splitting is
reversible (returning back to high temperature makes the lines to collapse) and
reproducible (many samples from different batches have been used). It is clear
that CO is responsible for the splitting of the EPR lines, but the physical
interpretation is not trivial. It is known that CO removes the center of
symmetry \cite{Nad2000} and reduces the group from P\={1}  to P1. The electronic
density is no more equivalent between the two TMTTF molecules, which leads to
two $^{13}$C NMR signals \cite{Chow2000}. However, the effect is different in
the case of EPR, where only one electron spin is shared by the two TMTTF
molecules, and thus only one signal is expected. Now, following the results of
Riera et Poilblanc \cite{Riera2001}, let's assume that the displacement of the
anions is not fully uniform and has a small modulation in space. In
Fig.\ref{fig:figlattice} we present two simple models. A and B are 2
representations of the unit cell in the CO state. In B the displacement of the
anions $X$ is a bit stronger than in A. Consequently, in B the electron-rich
TMTTF molecule is a bit richer than in A. Although the charge disproportion has
no direct effect on the spin, the counter-anion displacement has a similar
effect on the g-tensor. In triclinic symmetry, the the orientation of g-tensor
axes is not fixed by the symmetry and the displacement of $X$ induces a rotation
of the tensor principal axes\cite{Dutoit2015}. A similar effect has been
reported for other low-symmetry  systems \cite{VanRens1970,Pilbrow1979}. In the
configuration B the displacement is stronger than in A, which leads to a more
significant rotation of the g-tensor of the configuration B than in A.

\subsection{Exchange splitting}

The ability to resolve the two EPR lines coming from the A and B configurations
depends on relative value of the mismatch of the Zeeman energies for the two
non-equivalent spins, $\Delta g\mu_B H$, and on the Heisenberg exchange
interaction in the direction $i$, $k_B J_i$. When the magnetic field is smaller
than the exchange interaction, the signals from the two non equivalent sites are
merged into one line due to fast fluctuation, as predicted by the theory of
exchange narrowing \cite{Anderson1953,Kubo1954}. On the contrary, the lines from
the two non equivalent magnetic sites are split  \cite{Hennessy1973} in the case
of strong magnetic field:

\begin{equation}\label{eq:splitting}
\Delta g\mu_B H>k_B J_i.
\end{equation}

\begin{figure} \centering
	\includegraphics[width=0.8\columnwidth]{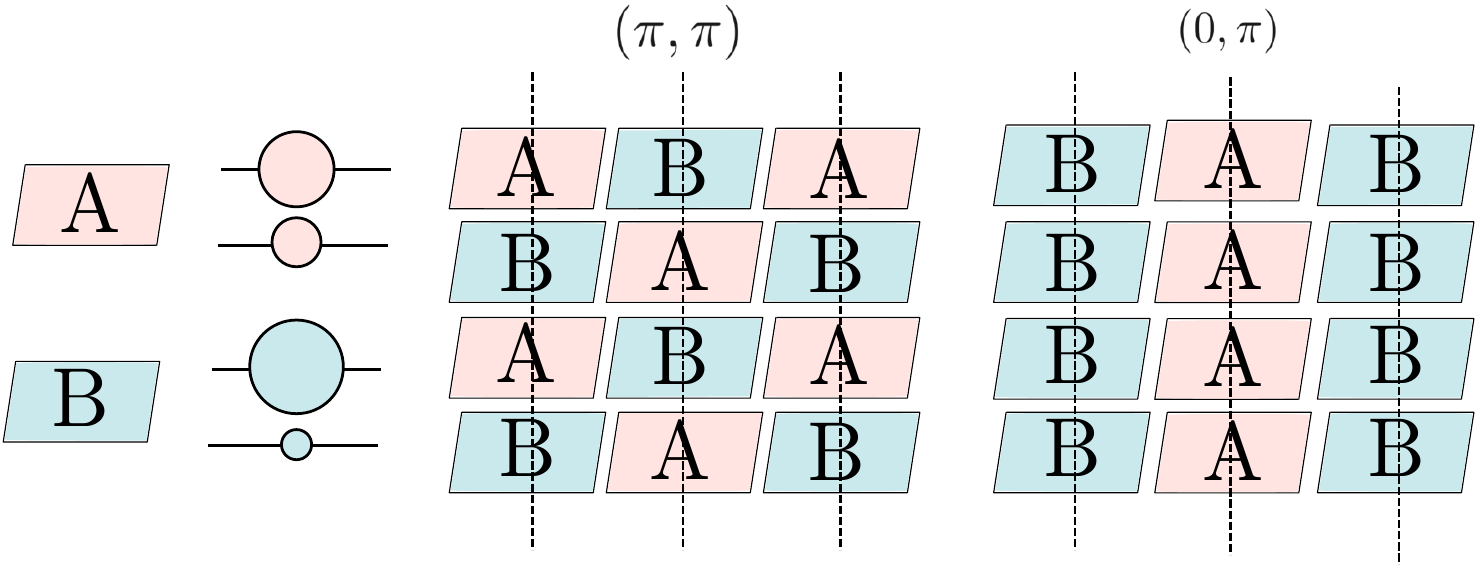}
	\caption{(color online). Schematic representation of CO superlattice inspired from the original
		work of Riera and Poilblanc \cite{Riera2001} . A and B are unit cells of
		(TMTTF)$_2$X in the CO configuration (ie : with charge density displacement ) but
		B has a stronger charge displacement than A. The vertical dashed lines represent
		the chain axis. The $(\pi,\pi)$ configuration alternates A and B in both
		directions whereas $(0,\pi)$ has a uniform A or B inside the chain but alternates
		between the chains.  } \label{fig:figlattice} 
	\end{figure}

Here $\Delta g$ is the difference of the g factors of the two non-equivalent
sites; it is a maximum at 45$^\circ$ between $b_m$ and $c_m$. This condition,
mathematically proven by P.W. Anderson \cite{W.Anderson1954},  was used to
directly estimate the interchain coupling in the quantum spin chain CuGeO$_3$
\cite{Nojiri1998}. The intrachain exchange interaction of (TMTTF)$_2$X
($\sim$400~K) is too large and all the chains in the configuration $(\pi,\pi)$
in Fig.\ref{fig:figlattice} should have the same resonance field (only one line
is expected). In the configuration  $(0,\pi)$, two kinds of chains exist : full
A and full B. Compared to Ref.\cite{Riera2001}, the configuration
$(\pi,\pi)$ is the scenario (c), $(\pi,0)$ is the scenario (d),  $(0,0)$ is  the
scenario (a) (and is the standard picture to explain the CO transition),
$(0,\pi)$ is not mentioned. Since we can resolve two lines, we are able to
decouple the signals from two different chains. Thus we are able to give an
upper limit for the smallest interchain coupling $J_c$.

\begin{table}
\begin{tabular}{|c|c|c|c|} \hline $X=$ & $\Delta g$ &
		\makecell{minimum frequency/(field) \\to resolve the splitting} & upper limit $J_c$ \\
		\hline AsF$_6$ & 12.10$^{-5}$ & 120~GHz (4.28~T) & 5G ($\sim$ 0.7~mK) \\ \hline SbF$_6$ &
		9.10$^{-5}$ & 240~GHz (8.56~T) & 8G ($\sim$ 1~mK) \\ \hline PF$_6$ & 7.10$^{-5}$ & 336~GHz (12~T) & 8G
		($\sim$ 1~mK) \\ \hline\end{tabular} 
	\caption{Estimation of the interchain coupling $J_c$. $\Delta g$ is the
	difference of the g factors of the two non-equivalent sites for
	$\theta=45^\circ$. The frequencies/fields correspond to the threshold where the
	splitting has been resolved. Using \eqref{eq:splitting} we estimate the upper
	limit of the interchain coupling.  }\label{table:1}
\end{table}

To our knowledge, this is the first experimental estimation of the interaction
between chains of (TMTTF)$_2X$ in the direction $c$ and this result has
remarkable implications for the magnetic properties.

It confirms the quasi absence of electronic correlation in the $c$ direction. Band
structure calculations have managed to estimate the transfer integral in the $a$
(chain axis) and $b$ directions \cite{Granier1988,Giovannetti2012} but only one
paper, so far, has reported a theoretical prediction in the $c$ direction 
$t_c\sim1$~meV, which leads in the tight binding model to $J_c =(t_c/t_a)^2J\sim
10$~mK) \cite{Grant1983} one order of magnitude higher than what we get
but not so far away from theoretical predictions using mean field
models. Indeed, the magnetic dimensionality has to be carefully considered.
Although the (TMTTF)$_2X$ family is a QQ1DAFH system, the large difference between
$J_b$ and $J_c$ makes some models fail. One remarkable
example is the  estimation of the interchain coupling $J'$ from the N\'{e}el
order temperature \cite{Irkhin2000,Yasuda2005}:

\begin{equation}\label{eq:intercoupling}
\left. J'=1.073 T_N^{1D} \middle/ \sqrt{\ln\left(\frac{2.6J}{T_N^{1D}}\right)+\frac{1}{2}\ln\ln\left(\frac{2.6J}{T_N^{1D}}\right)}\right.
\end{equation}

which gives for (TMTTF)$_2$SbF$_6$, $J'\sim 2$~K, three orders of magnitude
higher than the value we found. The reason comes from the assumptions made to
develop the standard model by Irkin and Katanin \cite{Irkhin2000} then Yasuda
\textit{et al.} \cite{,Yasuda2005} : The chain is coupled to the nearest neighbor chains
by the same interaction. This is not the case for (TMTTF)$_2$SbF$_6$. More
surprisingly, Yoshimi \textit{et al.} \cite{Yoshimi2012} have shown that in the
CO state the inter-site Coulomb repulsion could lead to an exchange interaction
of the same order in direction $a$ and $b$. However, this last result is not
confirmed  by our experimental results, since a QQ2DAFH should have a much higher
N\'{e}el temperature, even with the weak $J_c$ we report. Using the equation
connecting the N\'{e}el temperature and the interlayer coupling
$J'$\cite{,Yasuda2005}:

\begin{equation}\label{TN2D}
\left. T_N^{2D}=0.732\pi J  \middle/ \left(2.43-\ln\left(\frac{J'}{J}\right)- \ln\left(\frac{T_N^{2D}}{J}\right)  \right)\right.
\end{equation}
with $J=400$~K, $J'=1$~mK, we find $T_N^{2D}=54$~K, while $T_N^{1D}=3$~mK from eq.\eqref{eq:intercoupling}. In both cases, the model fails to describe
(TMTTF)$_2$SbF$_6$. Our results should help the extension of the 1D and 2D
models connecting $T_N$ and the exchange couplings \cite{Irkhin2000,Yasuda2005}
to the cases of intermediate dimensionality, $J_a>J_b\gg J_c$.

\subsection{Charge ordering and the superlattice in the $c$ direction.}

\begin{figure} \centering \includegraphics[width=0.8\columnwidth]{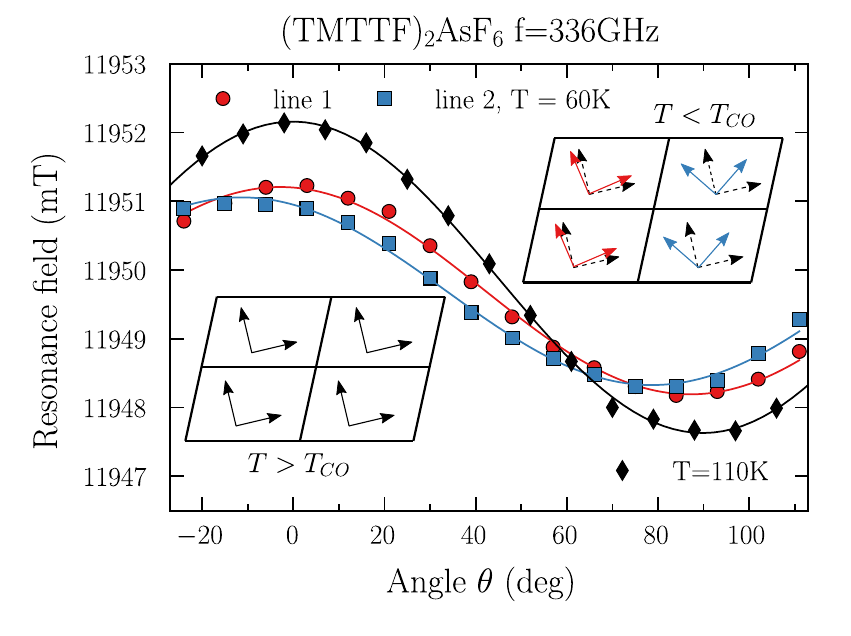}
	\caption{(color online). Resonance field of (TMTTF)$_2$AsF$_6$ recorded in the
		plane perpendicular to the chain axis. The black diamonds represent the line at
		$T=110$~K (above $T_{CO}$) while the red circles and the blue squares are the two
		lines observed at $T=60$~K (below $T_{CO}$). The lines are the best fits. The
		insets represent the g-tensor axes in the plane perpendicular to the chain axis.
		At $T>T_{CO}$ all the chains are equivalent. At $T<T_{CO}$ two inequivalent
		chains have different orientations of their g-tensors.  } \label{fig:Rotation}
\end{figure}

In order to understand the role of the CO transition on the appearance of a
magnetic superlattice in direction $c$, we performed EPR measurements for several
orientations in the plane perpendicular to the chain axis.  Fig.
\ref{fig:Rotation} shows the variation of the resonance field as a function of the
angle $\theta$ in (TMTTF)$_2$AsF$_6$ for $f=336$~GHz above (black diamonds) and below (red
circles and blue squares) $T_{CO}$. $\theta=0^\circ$ corresponds to the maximum of
the resonance field at $T=110$~K. At $T<T_{CO}$ the splitting of the EPR line is
accompanied by a rotation of the g-tensor. The rotation is not the same for the
two lines. In the case of line 1, we found a rotation through 3$^\circ$, whereas for
line 2 the angle of rotation is 11$^\circ$. The rotation of the principal axes of the
g-tensor was observed at low field in non-resolved EPR spectra and explained by
the displacement of the anions $X$ \cite{Dutoit2015}. The modulation of the
stain field in $c$ direction induces two kinds of anion displacements leading to
2 different rotations of the g-tensor which can be resolved at high 	field.

The relative rotation of the g-tensor ($\Delta \theta$) of two adjacent chains
was also assumed to explain the EPR line broadening for $T<T_{CO}$ at lower
frequencies \cite{Yasin2012}. However neither the rotation axis nor the
amplitude of the rotation is coherent with our results.  The reason is that the
authors used the anisotropic Zeeman effect (AZE) \cite{Pilawa1997} in order to
extract quantitative information from the EPR linewidth . Unfortunately, this
model needs the interchain coupling $J'$ and in the absence of a reliable value,
the authors estimated $J'$ using \cite{Yasuda2005}, which, as we have
shown previously, is irrelevant in the case of (TMTTF)$_2X$ salts.  Using
$J'=1.1$~K and an enhanced linewidth of 1.5~G they found a large relative
rotation of $\Delta\theta =\pm 32^\circ$ about the $b$ axis. Now let's approach
the problem from the other side by using our direct measurement of $\Delta
\theta=\pm 4^\circ$ about the $a$ axis which yields $\Delta g\approx 2.10^{-4}$,
and the experimental results obtained in W band from \cite{Yasin2012}
($\Delta H=1.5$~G and 	$H_0=3.36$~T). Applying the the AZE model :
\begin{equation}\label{eq:AZE}
J'=	\sqrt{\pi/8}H_0^2|\Delta g|^2/(g_e\Delta H)
\end{equation}
we find $J'\sim 0.9$~mK which is coherent with	our values of $J_c$ (Table \ref{table:1}).

\begin{figure} \centering \includegraphics[width=0.8\columnwidth]{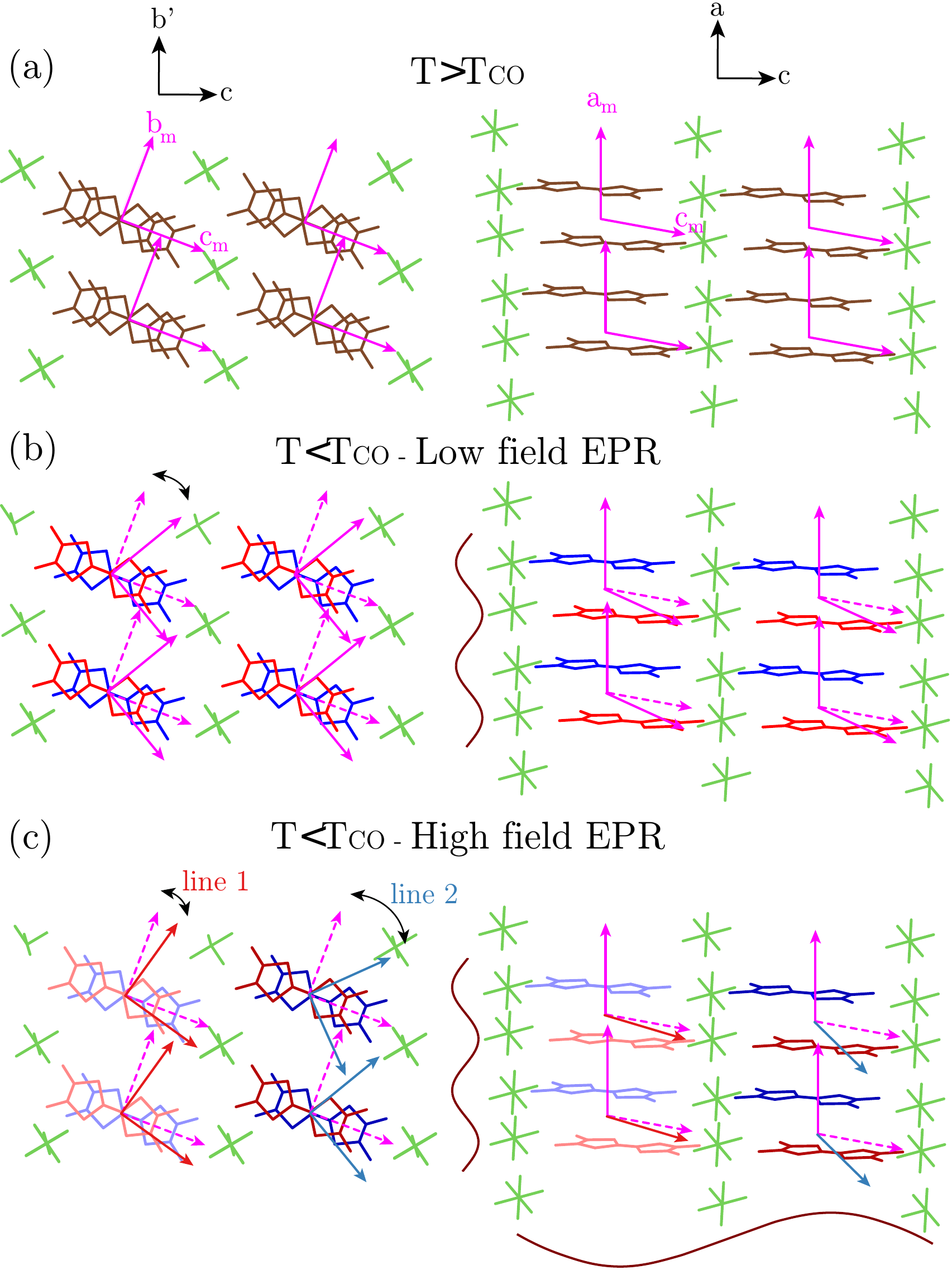}
	\caption{(color online). Schematic representation of the structural and
	magnetic properties of (TMTTF)$_2X$. The left column represents 4 unit cells in
	the plane perpendicular to the chain axis $a$ while the right side is 4 unit
	cells in the plane $ac$. (a) At $T>T_{CO}$, the charge density is equally
	distributed between the TMTTF molecules(brown). The principal axis of th
	g-tensor (pink) are oriented along the TMTTF molecular axes. All the unit cells
	are magnetically equivalent. (b) At $T<T_{CO}$ the charge density of the two
	TMTTF molecules is no more balanced and the g-tensor axes rotate about the $a$
	axis \cite{Dutoit2015}. (c) The high-field/high-frequency EPR
	experiments reveal that this rotation is not uniform due to modulation of the
	charge ordering. The charge unbalance is stronger in one stack (dark blue and
	red) than in the second (light blue and red) leading to two different rotations
	of the g-tensor axes and thus to two EPR lines. } \label{fig:CO}
\end{figure}

The figure \ref{fig:CO} demonstrates how EPR can probe the CO transition. Fig.
\ref{fig:CO}a: the TMTTF (brown) molecules are stacked in the $a$ direction and
form the chain axis. They are separated by the counter anions $X$ (PF$_6$,
AsF$_6$, SbF$_6$) (green). Due to the low triclinic symmetry of the system, the
crystallographic axes and the magnetic axes are different. For $T>T_{CO}$ the 2
TMTTF molecules are equivalent and the magnetic axes $c_m$ and $b_m$ coincide
with the symmetry axes of the TMTTF. At the CO phase transition (Fig.
\ref{fig:CO}b), the symmetry is reduced by removing the centers of inversion,
the counter anion $X$ cages are shifted and the charge balance on the two TMTTF
is broken. One TMTTF (blue) has a higher charge density than the second one
(red). Since the  g-tensor is sensitive to the electronic change
\cite{Dutoit2015}, it turns about the $a$ axis (as shown by low filed EPR
\cite{Furukawa2009,Dutoit2015}). The vertical sinusoidal curve represents the
change of the electronic density. Fig. \ref{fig:CO}c shows how the
high-field/high-frequency EPR revealed that the CO is not uniform and a CO in
the $c$ direction is also induced. In the chains with light (dark) red and blue
TMTTF the charge displacement is less (more) pronounced. As a consequence, the
rotation of the g-tensor is less (more) significant, resulting in the line 1
(line 2).

Finally, let's discuss why the superlattice in the $c$ direction was not
detected by other techniques. In the introduction to this paper we have
presented several techniques which have permitted he observation of the CO.
However, only few of them such as $^{13}$C NMR  or neutron scattering, have the
spectral or spatial resolution to detect the interchain CO modulation. In the
case of  $^{13}$C NMR, the main CO was detected by observing the splitting of 
$^{13}$C NMR lines but, as seen in Fig. 2 and 4 of Ref \cite{Chow2000},
this splitting is very small. It is not surprising that a modulation of the CO
would lead to a second separation of each line too tiny to be resolved. The
measurement of the displacement of the counter anions responsible for the CO was
a challenging neutron scattering experiment \cite{Foury-Leylekian2010}. The
modulation of the displacement would just be included in the error bars. In our
case the observation of the CO modulation was possible because of : (\textit{i})
The low P$_1$ symmetry which allows the rotation of the g-tensor axes. In higher
symmetry, this effect would not exist. (\textit{ii}) The very small linewidth of
the organic salts ($\Delta H\sim 1$~G) allows to resolve the two EPR lines. 
(\textit{iii}) The two EPR lines are not collapsed by the interchain exchange
interaction because electronic correlations in the $c$ direction were nearly
absent and the magnetic field was high (exchange splitting
regime\cite{W.Anderson1954}).

In conclusion, the CO transition observed in (TMTTF)$_2X$  ($X=$SbF$_6$,
AsF$_6$, PF$_6$) is accompanied by a displacement of the counter anions $X$.
This displacement was observed by neutron \cite{Foury-Leylekian2010}, NMR
\cite{Chow2000} and dilatometry \cite{Monceau2001}  but was considered uniform
\cite{DeSouza2008} all over the crystal. Using high field/frequency EPR, we have
seen the signal of two non-equivalent spin chains. We have interpreted this
result as a modulation of the displacement of $X$ in the direction $c$ leading
to two orientations of the g-tensor and consequently to two EPR lines.

The line splitting allows us to estimate the exchange coupling in the $c$
direction ($J_c\sim 1$~mK). Using this value, we have shown that neither the 1D
model (equation \eqref{eq:intercoupling}) nor the 2D one ( equation
\eqref{TN2D}) can describe (TMTTF)$_2$SbF$_6$. We believe that this result
should stimulate the development of a theory linking $T_N$ and the inter chain
couplings in the case $J_a>J_b\gg J_c$.

\section{Acknowledgement.}

We acknowledge M. Dressel for providing the sample.  We thank M. Kuzmin and S. Todo for stimulating discussions. This work was supported by
NSF Grant No. DMR-1206267, CNRS-PICS CoDyLow and CNRS's infrastructure of
research  RENARD (FR3443) for EPR facilities. The NHMFL is supported by the
Cooperative Agreement Grant No. DMR-1157490 and the State of Florida.

 \bibliographystyle{achemso}
\providecommand{\latin}[1]{#1}
\makeatletter
\providecommand{\doi}
{\begingroup\let\do\@makeother\dospecials
	\catcode`\{=1 \catcode`\}=2 \doi@aux}
\providecommand{\doi@aux}[1]{\endgroup\texttt{#1}}
\makeatother
\providecommand*\mcitethebibliography{\thebibliography}
\csname @ifundefined\endcsname{endmcitethebibliography}
{\let\endmcitethebibliography\endthebibliography}{}


\end{document}